\documentclass[prd,showpacs,preprintnumbers,amsmath,amssymb]{revtex4}

\usepackage{graphicx}
\usepackage{dcolumn}
\usepackage{bm}

\def\be{\begin{eqnarray}}
\def\ee{\end{eqnarray}}
\def\bea{\begin{eqnarray}}
\def\eea{\end{eqnarray}}
\def\kT{{\bf k}_\perp}
\def\kp{{\bf k}_\perp}

\def\yT{{\bf y}_\perp}
\def\xT{{\bf x}_\perp}

\def\bT{{\bf b}_\perp}
\def\bp{{\bf b}_\perp}

\def\0T{{\bf 0}_\perp}
\begin{document}


\title{Transverse Deformation of Parton Distributions and Transversity
Decomposition of Angular Momentum}

\author{Matthias Burkardt}
 \affiliation{Department of Physics, New Mexico State University,
Las Cruces, NM 88003-0001, U.S.A.}
\date{\today}

\begin{abstract}
Impact parameter dependent parton distributions are transversely
distorted when one considers transversely polarized nucleons and/or
quarks. This provides a physical mechanism for the T-odd
Sivers effect in semi-inclusive deep-inelastic 
scattering. The transverse distortion can also be
connected with Ji's quark angular momentum relation. 
The distortion of chirally odd impact parameter dependent 
parton distributions is related to chirally odd GPDs. This result 
is used to provide a decomposition of the quark angular momentum
w.r.t. quarks of definite transversity. Chirally odd GPDs can thus
be used to determine the correlation between quark spin and quark
angular momentum in unpolarized nucleons. Based on the transverse 
distortion, we also suggest a qualitative connection between
chirally odd GPDs and the Boer-Mulders effect.
\end{abstract}

\maketitle
\narrowtext
\section{Introduction}
During the last few years, important breakthroughs have been made
in our understanding of T-odd single-spin asymmetries (SSA) in 
semi-inclusive deep-inelastic scattering (SIDIS) 
\cite{ansel,mulders}. In a seminal, paper Brodsky, Hwang and
Schmidt \cite{BHS}, 
provided a simple model calculation in which the 
interference of final state interaction (FSI) phases 
between different partial waves gave rise to a nontrivial
Sivers effect \cite{sivers}. This calculation clearly demonstrated
that T-odd distributions can also survive in the Bjorken limit in QCD.
Following this work, the connection
between these FSI phases and the Wilson line
gauge links in gauge invariantly defined unintegrated parton
densities was recognized \cite{collins,ji}. 
This also led to the prediction
that, up to a sign, the Sivers functions in SIDIS and polarized
Drell-Yan are the same \cite{collins}.
Soon later, an intuitive connection between the sign of the
Sivers effect and the transverse distortion of impact parameter
dependent parton distributions in transversely polarized targets 
was proposed \cite{mb1}.
This connection also explained the similarity between the
light-cone overlap integrals relevant for the Sivers effect and
for the anomalous magnetic moment \cite{BHS2}. 

Generalized parton distributions (GPDs) 
provide a decomposition of form 
factors at a given value of $t$, w.r.t. the average 
momentum fraction $x = \frac{1}{2}\left(x_i+x_f\right)$ 
of the active quark 
\bea
\int dx H_q(x,\xi,t) &=& F^q_1(t)
\quad \quad \int dx \tilde{H}_q(x,\xi,t) = G^q_A(t)
\label{GPD1}\nonumber\\
\int dx E_q(x,\xi,t) &=& F^q_2(t)\quad \quad 
\int dx \tilde{E}_q(x,\xi,t) = G^q_P(t)
\label{GPD} ,
\eea
where $x_i$ and $x_f$ are the momentum fractions of the quark
before and after the momentum transfer. $2\xi=x_f-x_i$ represents
their difference. For recent reviews, with more precise definitions
and a detailed discussion 
of their early history, the reader is referred to Refs. 
\cite{GPD1,GPD2,GPD3,GPD4}.
$F^q_1(t)$, $F^q_2(t)$, $G^q_A(t)$, and $G^q_P(t)$ are the Dirac,
Pauli, axial, and pseudoscalar formfactors, respectively.
Note that the measurement of the quark momentum fraction $x$ 
singles out one space direction (the direction of the momentum).
Therefore, it makes a difference whether the momentum transfer
is parallel, or perpendicular to this momentum. The GPDs must 
therefore depend on an additional variable which characterizes
the direction of the momentum transfer relative to the momentum
of the active quark. Usually, one parameterizes this dependence
through the dimensionless variable $\xi$. Throughout this work we
will focus on the limiting case $\xi=0$, where GPDs can be 
interpreted as the Fourier transform of the distribution
of partons in the transverse plane 
(see Refs. \cite{mbGPD,gpde,soper,DiehlEPJ,JiWig} and references 
therein). 

The impact parameter dependent distributions are defined
as follows. First one introduces nucleon states which are 
localized in
transverse position space at ${\bf R}_\perp$ (they are eigenstates
of the transverse center of momentum with eigenvalue ${\bf R}_\perp$)
\be
\left| p^+,{\bf R}_\perp, \lambda\right\rangle
= {\cal N} \int \frac{d^2 {\bf p}_\perp}{(2\pi)^2}
e^{-i{\bf p}_\perp {\bf R}_\perp} 
\left| p^+,{\bf p}_\perp, \lambda\right\rangle
\ee
where ${\cal N}$ is some normalization factor. 
In these localized states, the  impact parameter dependent 
distributions are then defined as the familiar light-cone 
correlations. For further details, see Refs. \cite{gpde,DiehlEPJ}.

For example, for the impact parameter dependent distribution 
$q(x,\bp)$
of unpolarized quarks in an unpolarized target one finds
for the distribution of quarks with momentum fraction $x$
\be
q(x,\bp) = {\cal H}(x,\bp) \equiv \int 
\frac{d^2 {\bf \Delta}_\perp}{(2\pi)^2} e^{-i\bp {\bf \Delta}_\perp}
H(x,0,-{\bf \Delta}_\perp^2),
\label{q1}
\ee
where $\bp$ is the transverse distance from the active quark to
the transverse center of momentum
\be
{\bf R}_\perp = \sum_i x_i {\bf r}_{\perp,i}
\label{Rperp}.
\ee
and ${\bf \Delta}_\perp = {\bf p}_\perp^\prime -{\bf p}_\perp$.
The transverse center of momentum is the analog of the nonrelativistic
center of mass and the sum in Eq. (\ref{Rperp}) extends over
both quarks and gluons. The $x_i$ are the momentum fractions
of each parton which play the same role that the mass fraction plays
in nonrelativistic physics. One of the remarkable features of Eq.
(\ref{q1}) is that there are no relativistic corrections to the
interpretation of GPDs (at $\xi=0$) as Fourier transforms of parton
distributions in impact parameter space \cite{mb1}. 
This is due to the
presence of a Galilean subgroup of transverse boosts in the 
light front formulation of relativistic dynamics \cite{Dirac}. 
Another remarkable feature is that the impact parameter dependent
parton distributions obtained via Eq. (\ref{q1}) have a probabilistic
interpretation and satisfy corresponding positivity constraints
\cite{positivity}.

For the distribution $q_X(x,\bp)$ of unpolarized quarks
in a nucleon state that is a superposition of positive and negative
(light-cone) helicity states
\be
\left|X\right\rangle \equiv \frac{1}{\sqrt{2}}\left[
\left|p^+,{\bf R}_\perp, +\right\rangle
+\left|p^+,{\bf R}_\perp, -\right\rangle\right]
\label{X}
\ee
one finds (for details see Ref. \cite{gpde}, where a detailed
definition of these distributions is provided)
\be
q_X(x,\bp) = {\cal H}(x,\bp)-\frac{1}{2m}
\frac{\partial}{\partial b_y} {\cal E}(x,\bp)
\label{q2}
\ee
where
\be
{\cal E}(x,\bp) \equiv \int 
\frac{d^2 {\bf \Delta}_\perp}{(2\pi)^2} e^{-i\bp {\bf \Delta}_\perp}
E^q(x,0,-{\bf \Delta}_\perp^2).
\label{q3}
\ee
Up to relativistic corrections, due to the transverse
localization of the wave packet (these corrections will be discussed
below in connection with the angular momentum relation), 
this state can be interpreted as a transversely
polarized target and Eq. (\ref{q3}) predicts that the impact
parameter dependent parton distributions give rise to
a transverse flavor dipole moment in a transverse polarized target.
The average magnitude of this distortion is normalized to the
anomalous magnetic moment contribution from that quark flavor
\cite{gpde}. The physical origin of this distortion is the fact that
the virtual photon in DIS couples only to the $j^+=j^0+j^3$ component
of the quark density in the Bjorken limit. For quarks with
nonvanishing orbital angular momentum, the $j^3$ component of the
quark current has a left-right asymmetry due to the orbital motion
\cite{JiPRL2}. 

The transverse distortion of the parton distributions
exhibited in Eq. (\ref{q2}), in combination with an attractive
final state interaction, has been suggested as a simple explanation
for the Sivers effect in QCD \cite{mb2,mb3}. 
Since the sign of $E^q$ can be 
related to the contribution from quark flavor $q$ to the
anomalous magnetic moment of nucleons, Eq. (\ref{q2}) has been
the basis for a prediction of the signs of the Sivers effect
for $u$ and $d$ quarks \cite{mb2,mb3}, 
which have been confirmed by the HERMES
collaboration \cite{hermes}.

In this work we will first discuss the connection between the
transverse distortion of GPDs and Ji's quark 
angular momentum relation \cite{JiPRL}.
This will allow us to draw a link between chirally odd GPDs and
the correlation between the angular momentum and spin of the quarks.
We also propose a simple
explanation for the Boer-Mulders effect \cite{BM}, 
where the asymmetry
arises from the transverse distortion of chirally odd GPDs.

\section{Transverse Component of the Angular Momentum}
In this section, we discuss the connection between the transverse 
distortion of quark distribution in impact parameter space and
Ji's quark angular momentum relation
\be
J^i= \frac{1}{2}\varepsilon^{ijk} \int d^3x M^{0jk}.
\label{M0jk}
\ee
The angular momentum density
$M^{\alpha \mu \nu} = T^{\alpha \nu}x^\mu - T^{\alpha \mu} x^\nu$ is 
expressed in terms of the energy momentum tensor $T^{\mu \nu}$.

Since the angular momentum operator is expressed in terms of the
position space moments of the energy momentum
tensor, it is possible to relate $J_q$ to the form factor of
the energy momentum tensor \cite{JiPRL}
\be
\left\langle p^\prime \left| T_q^{\mu \nu} \right|p\right\rangle
= \bar{u}(p^\prime)\left[  A_q(\Delta^2) \gamma^\mu \bar{p}^\nu 
+B_q(\Delta^2) \frac{i\sigma^{\nu \alpha}}{2M}\bar{p}^\mu
\Delta_\alpha + C_q(\Delta^2)\frac{\left( \Delta^\mu \Delta^\nu
-g^{\mu \nu} \Delta^2 \right)}{M}
+ \bar{C}_q(\Delta^2) g^{\mu \nu}M\right] u(p),
\label{T}
\ee
where symmetrization of the indices $\mu$ and $\nu$ is implicit
and $2\bar{p}^\mu = p^\mu + p^{\prime \mu}$. The label 
$q$ distinguishes the form factors of the
different quark flavors or the glue. The angular momentum relation
obtained from Eqs. (\ref{M0jk}) and (\ref{T}) reads \cite{JiPRL}
\be
\left\langle J^i_q \right \rangle = S^i\left[ A_q(0) + B_q(0)\right],
\label{JiPRL}
\ee
where $S^i$ is the nucleon spin. Further details can be found
in Ref. \cite{JiPRL}.

For the relation between the transverse deformation of impact 
parameter dependent parton distributions and the angular
momentum, we now concentrate on the form factor of the
`good' component of the energy momentum tensor
\be
\left\langle p^\prime |T^{++}_q(0)|p\right\rangle 
= \bar{u}(p^\prime)\left[
A_q(-{\bf \Delta}_\perp^2)\gamma^+p^+ + B_q(-{\bf \Delta}_\perp^2)
p^+\frac{i\sigma^{+i}\Delta_i}{2M}\right]u(p) .
\label{T++}
\ee
In this work, we are mainly interested in the spin component
perpendicular to the light-cone direction, which is
sensitive to longitudinal boosts.
It is thus important to specify the frame.
Here and in the following we only consider the special case 
$p^+=p^{\prime +}=M$, i.e. our results apply to the rest frame
of the target. 

Application of Eq. (\ref{T++}) to 
a delocalized wave packet $\left|\psi\right\rangle$ of a transversely
polarized nucleon with transverse spin $S^j$ yields
\be
\left\langle \psi \left|\int d^2{\bf b}_\perp b_i 
T_q^{++}(\bp) \right|\psi\right\rangle
= {\cal N} \varepsilon_{ij}
S^j p^+ \left[A_q(0)+B_q(0)\right],
\label{T++AB}
\ee
where ${\cal N}$ is a normalization factor depending on the wave 
packet. Ideally, we would
like to take the expectation value in Eq. (\ref{T++AB}) in
plane wave state with ${\vec p}=0$, but this leads to
ill defined expressions when $b\rightarrow \infty$. In order to 
regularize these expressions, we thus use
$\left|\psi\right\rangle = \int d^3k\, \psi({\vec k}) \left|{\vec k},
{\vec S}\right\rangle$, where $\left|{\vec k}, {\vec S}
\right\rangle$ are spin 
eigenstates and imagine taking the limit where
$\psi({\vec k})$ is nonzero only for ${\vec k}=0$ in the end of the 
calculation. In the light-cone analysis, we have in mind taking the
limit $k^z=0$ immediately and choosing a dependence
on ${\bf k}_\perp$ that is axially symmetric (depends only on 
$|{\bf k}_\perp|^2$). For the discussion in a general frame, 
$\psi({\vec k})=\psi( {\vec k}^2)$. the actual shape of the 
wave functions entering these wave packets is irrelevant after
the limit ${\vec k}\rightarrow 0$ has been taken.
A comparison betwwen Eq. (\ref{T++AB}) and 
Eq. (\ref{JiPRL}) shows that the expectation value of the
angular momentum of the quarks $\left\langle J_q^i \right\rangle$ 
in a transversely polarized delocalized state can be related
to the transverse center of momentum of the quarks in the same state.
This observation provides a physical
explanation for Ji's result
linking $J_q$ and the GPDs $H_q(x,0,0)$ and $E_q(x,0,0)$.

Moreover, in light-cone gauge $A^+=0$, $T^{++}$ contains no 
interactions
between the fields and it is natural to decompose
\be
T^{++} = T^{++}_q + T^{++}_g= i q_+^\dagger \partial^+q_+
+ Tr\left(\partial^+ {\vec {\bf A}}_\perp^2\right),
\ee
where $q_+ = \frac{1}{2} \gamma^-\gamma^+q$ is the ``good'' 
component. This provides a parton model interpretation
for Ji's quark angular momentum relation. Upon
switching to a mixed representation (momentum/
position space representation for the longitudinal/transverse
coordinate respectively) and express the transverse shift of the 
center
of momentum for a particular quark flavor in terms of the
impact parameter dependent parton distributions, 
yielding
\be
\label{J}
\left\langle \psi \left| J^i_q \right|\psi\right\rangle = 
\varepsilon^{ij}
M \int dx \int d^2{\bf b}_\perp 
q_\psi(x,\bp) x 
b^j,
\ee
where $q_\psi(x,\bp)$ is the impact parameter dependent
parton distribution evaluated in the state $\psi$.

We should emphasize that it is crucial for this argument that
we work with a delocalized state which is centered around the origin.
As a counter example, when
As an application of Eq. (\ref{J}), we now insert $q_X(x,\bp)$ 
for the ``transversely polarized'' state 
above (\ref{q2}), yielding
\be
\left\langle X \left| J^x_q \right| X\right\rangle =
\frac{1}{2} \int dx E_q(x,0,0)x,
\ee
which is obviously only part of Eq. (\ref{JiPRL}).
In order to better understand the connection between 
impact parameter dependent PDFs and the angular momentum of the
quarks, we now investigate the origin of this discrepancy further.
For this purpose we note that the state
$\left|X\right\rangle$ (\ref{X})
is localized in impact parameter space, and its momentum space
wave packet contains an integral over transverse momentum.
However, for states with a nonzero transverse momentum, 
the light-front
helicity eigenstates and the rest frame spin eigenstates are not 
the same. Indeed, already for $|\kp|\ll M$ one finds \cite{melosh}
\bea
\left|k,+\right\rangle_I &=& \left|k,+\right\rangle_F -
\frac{k_R}{2M}\left|k,-\right\rangle_F
\\
\left|k,-\right\rangle_I &=& \left|k,-\right\rangle_F +
\frac{k_L}{2M}\left|k,+\right\rangle_F
\eea
where $k_R=k_1+ik_2$ and $k_L=k_1-ik_2$. The subscripts ``I'' and
``F'' refer to the spin eigenstates in the instant as well as
front form of dynamics respectively \cite{Dirac}. Explicit 
representations for instant and front form spinors and the
transformation relating them can also be found in the appendix 
of Ref. \cite{GPD2}.
As a consequence of this ``Melosh rotation'', one should only identify
the state $\left|X\right\rangle$ with a state that is transversely
polarized in its rest frame up to relativistic corrections.

This well-known
result has important consequences if we consider a delocalized
state that is polarized in the $+\hat{x}$ direction in the rest 
frame
\bea
\left| \psi^{+\hat{x}}_I\right\rangle
&\equiv& \int d^2 \kT \psi^{+\hat{x}}_I(\kT) \left|\kT,+\hat{x}
\right\rangle_I
\\
&=&\int d^2 \kT \left[ 
\psi^{+\hat{x}}_F(\kT) \left|\kT, +\hat{x}\right\rangle_F+
\psi^{-\hat{x}}_F(\kT) \left|\kT,-\hat{x}\right\rangle_F\right]
\eea
with
\bea
\psi^{+\hat{x}}_F(\kT)&=& \left(1-i\frac{k_2}{2m}\right) 
\psi^{+\hat{x}}_I(\kT)\\
\psi^{-\hat{x}}_F(\kT)&=& \frac{k_1}{2M} 
\psi^{+\hat{x}}_I(\kT)\nonumber .
\label{eq:shift}
\eea
For the wave packet $\psi^{+\hat{x}}_I(\kT)$ in the rest frame we
have in mind an axially symmetric function that describes
a state that is delocalized in transverse position space, but centered
around the origin. The longitudinal momentum 
$k^+=k^0+k^3\approx M$ 
is kept fixed.

It is fallacious to believe that this  effect is negligible in the
limiting case of a delocalized wave packet. Indeed, to leading order
in $\frac{1}{M}$ (higher orders in $\frac{1}{M}$ involve
additional powers of the size $R$ of the wave packet in the 
denominator and are suppressed for a large wave packet),
the factor $\left(1-i\frac{k_2}{2M}\right)$ 
implies that the corresponding position space wave packet
in the front form is shifted sideways by half a Compton wavelength
\be
\tilde{\psi}^{+\hat{x}}_F(\bp)&=& 
\tilde{\psi}^{+\hat{x}}_I\left(\bp -\frac{1}{2M}\hat{{\bf y}}\right).
\ee
To leading order in $\frac{1}{M}$ there is no significant effect
from $\psi^{-\hat{x}}_F$, since all contributions to the
center of momentum are proportional to 
$\left| \tilde{\psi}^{-\hat{x}}_F(\bp)\right|^2 \sim 
\frac{1}{M^2}$, i.e. for dimensional reasons they must also be 
or order $\frac{1}{R}$.
The sideways shift implies that a large 
axially symmetric wave packet for a spin $\frac{1}{2}$ particle,
polarized in the $+\hat{x}$ direction, that is centered around the
origin in the rest frame corresponds again to a particle polarized
in the $+\hat{x}$ direction in the front form, but now the
wave packet is centered around $\bp=+\frac{1}{2M}\hat{{\bf y}}$.
For a particle that is polarized in the $-\hat{x}$ direction the
shift is in the opposite direction. 

This phenomenon has a number of applications. First
it explains  how an elementary Dirac particle, for which
\be
q_X(x,\bp)=\delta(x-1)\delta(\bp)
\label{bare}
\ee
can yield a nontrivial result for its total angular momentum
from Eq. (\ref{J}): For a state that is polarized in the
$+\hat{x}$ direction, and which in the instant form is described
by a wave packet $\tilde{\psi}_{+\hat{x}}(\bp)$, the corresponding
front form wave packet is centered around $\bp=+\frac{1}{2M}\hat{y}$.
In general, the distribution of partons in a wave packet 
$q_\psi(x,\bp)$ is obtained
by convoluting the intrinsic distribution $q(x,\bp)$ 
(relative to the center of momentum) with the distribution
$|\psi(\bp)|^2$ resulting from the wave packet. For our example
of an elementary Dirac particle, this implies
\be
q_\psi(x,\bp) = \int d^2{\bf r}_\perp 
\left|\tilde{\psi}^{+\hat{x}}_F( {\bf r}_\perp )\right|^2 
q(x,\bp+{\bf r}_\perp) =\delta(x-1) \left|\tilde{\psi}^{+\hat{x}}_F( {\bf b}_\perp )\right|^2 = \delta(x-1)
\left|\tilde{\psi}^{+\hat{x}}_I( {\bf b}_\perp -
\frac{1}{2M}\hat{\bf y})\right|^2
\ee 
plus corrections that are negligible for a large wave packet.
upon integrating over $\bp$, one easily finds
\be
M\int d^2\bp q_\psi(x,\bp) \bp = \frac{1}{2}\delta(x-1)
\ee
and therefore 
\be
\left\langle J_q^x\right\rangle = M\int dx 
\int d^2\bp q_\psi(x,\bp) b^y = \frac{1}{2}.
\ee
From the derivation it should be clear that the sideways shift
by $b^y = \frac{1}{2M}$ is essential for this result.

The second application is to a spin $
\frac{1}{2}$ particle polarized in the $+\hat{x}$ direction
with a nontrivial intrinsic distribution $q_X(x,\bp)={\cal H}(x,\bp)
-\frac{1}{2M}\frac{\partial}{\partial b_y}{\cal E}(x,\bp)$.
In this case
\be
q_\psi(x,\bp) = \int d^2{\bf r}_\perp 
\left|\tilde{\psi}^{+\hat{x}}_F( {\bf r}_\perp )\right|^2 
q(x,\bp+{\bf r}_\perp)
\ee
and the resulting transverse flavor dipole moment receives contributions
both from the wave packet as well as from the intrinsic distortion.
After an appropriate shift of variables one easily finds
\bea
\int d^2\bp q_\psi(x,\bp) \bp &=& 
\int d^2\bp q(x,\bp)
\int d^2{\bf r}_\perp 
\left|\tilde{\psi}^{+\hat{x}}_F( {\bf r}_\perp )\right|^2 
{\bf r}_\perp
+ \int d^2{\bf r}_\perp 
\left|\tilde{\psi}^{+\hat{x}}_F( {\bf r}_\perp )\right|^2 
\int d^2\bp q(x,\bp) \bp \\
&=& \frac{1}{2M}
\left[H(x,0,0) + E(x,0,0)\right].\nonumber
\eea
From the point of view of impact parameter dependent PDFs,
the $H(x,0,0)$ contribution in Ji's relation is thus
due to the sideways shift of the wave packet in the transition from
the instant form to the front form description
\be
\left\langle J_q^x\right\rangle = M\int dx 
\int d^2\bp q_\psi(x,\bp) \bp = \frac{1}{2}\int dx 
\left[H(x,0,0) + E(x,0,0)\right]x.\label{Jx}
\ee
Although, due to rotational invariance, the final result holds
for any component of ${\vec J}_q$, in the context of impact parameter
dependent parton distributions the Ji relation
naturally emerges
as a relation for $J_q^\perp$ . In particular it appears natural
to identify the integrand of Eq. (\ref{Jx}) 
$\frac{1}{2}\left[H(x,0,0) + E(x,0,0)\right]x$ with a momentum
decomposition of the transverse component of the quark angular momentum
in a transverse polarized target. The term containing $E(x,0,0)$
arises from the transverse deformation of GPDs in the center of 
momentum
frame, while the term containing $H(x,0,0)$ in Ji's relation
 arises
from an overall transverse shift when going from transverse polarized
nucleons in the instant form (rest frame) to the front form
(infinite momentum frame).

\label{sec:LC}
\section{Chirally Odd GPDs}
Similar to the chirally even case, chirally odd GPDs 
are defined as non-forward matrix elements of light-like
correlation functions of the tensor charge
\bea
p^+\int \frac{dz^-}{2\pi} e^{ixp^+z^-}
\left\langle p^\prime \right|
\bar{q}\left(-\frac{z}{2}\right) \sigma^{+j}\gamma_5
q\left(\frac{z}{2}\right)\left|p\right\rangle
&=& H_T(x,\xi,t) \bar{u} \sigma^{+j}\gamma_5u
+\tilde{H}_T(x,\xi,t)\varepsilon^{+j\alpha\beta}
\bar{u}\frac{\Delta_\alpha p_\beta}{M^2}u
\\
& &+{E}_T(x,\xi,t)\varepsilon^{+j\alpha\beta}
\bar{u}\frac{\Delta_\alpha \gamma_\beta}{2M}u
+\tilde{E}_T(x,\xi,t)\varepsilon^{+j\alpha\beta}
\bar{u}\frac{p_\alpha \gamma_\beta}{M}u.\nonumber
\eea
\begin{figure}
\unitlength0.90cm
\begin{picture}(15,7.0)(-9.0,0.5)
\includegraphics{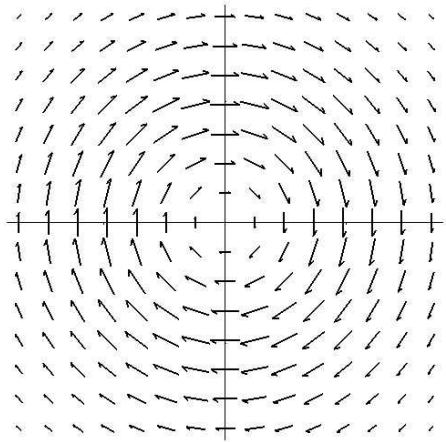}
\end{picture}
\caption{Distribution of transversity in impact parameter space
for a simple model (\ref{model}).}
\label{wirbel}
\end{figure}        
The connection between GPDs at $\xi=0$ and parton distributions
in impact parameter space has recently been extended to the chirally
odd sector \cite{DH}. Quarks $q(x,\bT,{\bf s})$ 
with transverse polarization 
${\bf s}= (\cos \chi,\sin \chi)$
are projected out by the operator $\frac{1}{2}\bar{q}\left[
\gamma^+-s^j i\sigma^{+j}\gamma_5\right]q$. Even for an unpolarized
target, the transversity density $\delta^i q(x,\bp)$, obtained from
\be
s_i \delta^i q(x,\bp) =
q(x,\bT,{\bf s})-q(x,\bT,-{\bf s}),
\ee
can be nonzero. Indeed, in Ref. \cite{DH} it is shown that
\be
\label{Diehl}
\delta^i q(x,\bp) \equiv - \frac{\varepsilon^{ij}}{2M} 
\frac{\partial}{\partial b_j} \left[
2\tilde{\cal H}_T(x,\bp) +{\cal E}_T(x,\bp)\right],
\ee
and
\bea
\tilde{{\cal H}}_T(x,\bp) 
&\equiv& \int \frac{d^2{\bf \Delta}_\perp}{(2\pi)^2}
e^{-i\bp\cdot{\bf \Delta}_\perp} \tilde{H}_T
(x,0,-{\bf \Delta}_\perp^2)
\\
{\cal E}_T(x,\bp) &\equiv& \int \frac{d^2{\bf \Delta}_\perp}{(2\pi)^2}
e^{-i\bp\cdot{\bf \Delta}_\perp} E_T(x,0,-{\bf \Delta}_\perp^2)
\nonumber
\eea
Eq. (\ref{Diehl}) exhibits a nontrivial flavor dipole moment
perpendicular to the quark spin
\be
\int d^2\bp \delta^i q(x,\bp) b_j = \frac{ \varepsilon^{ij}}{2M} 
\left[2\tilde{H}_T(x,0,0)+E_T(x,0,0)\right].
\label{spindipole}
\ee
The resulting effect is best illustrated in a simple model
[Fig. \ref{wirbel}]
\be
2\tilde{\cal H}_T(x,\bp) +{\cal E}_T(x,\bp) \propto \exp(-\bp^2).
\label{model}
\ee
Physically, the nonvanishing transversity
density in an unpolarized target is due to spin-orbit correlations
in the quark wave functions: if the quarks have orbital angular
momentum then their $\gamma^+$-density is enhanced on one side,
i.e. their distribution appears shifted sideways \cite{gpde,JiPRL2}. 
For
unpolarized nucleons all orientations of the orbital angular momentum
are equally likely and therefore the unpolarized quark distribution
is axially symmetric. However, if there is a correlation between
the orientation of the quark spin and the angular momentum then
quarks of a certain orientation will be shifted towards one side, 
while those with a different orientation will be shifted towards
a different side. 
                                                  
\section{Transversity Decomposition of the Angular Momentum}
In the discussion about the physical origin of the transversity
distribution in an unpolarized target, we hinted already at 
a connection between the linear combination $2\tilde{H}_T +E_T$ of
chirally odd GPDs
on the one hand and the correlation of quark spin and angular momentum on
the other hand.
In order to quantify this phenomenon, we are now considering a
decomposition of the quark angular momentum with respect to
quark transversity.
Such a decomposition is possible since $T^{++}_q$, whose form factors
enter Ji's quark angular momentum relation \cite{JiPRL} does not mix 
quark transversity states. Indeed, if we denote positive and
negative helicity states with $q_\rightarrow$ and $q_\leftarrow$
respectively, one finds
\bea
T^{++}_q =i\bar{q}\gamma^+
\stackrel{\leftrightarrow}{D^+}q &=& 
i\bar{q}_{\rightarrow}\gamma^+
\stackrel{\leftrightarrow}{D^+} q_\rightarrow + 
i\bar{q}_{\leftarrow}\gamma^+
\stackrel{\leftrightarrow}{D^+}q_\leftarrow\\
&=& \frac{i}{2}\left(\bar{q}_{\rightarrow}+
\bar{q}_{\leftarrow}\right) 
\gamma^+
\stackrel{\leftrightarrow}{D^+} 
\left(q_\rightarrow +q_\leftarrow\right)
+
\frac{1}{2}\left(\bar{q}_{\rightarrow}-\bar{q}_{\leftarrow}\right) 
\gamma^+
\stackrel{\leftrightarrow}{D^+} 
\left(q_\rightarrow-q_\leftarrow\right)
=T^{++}_{q,+\hat{x}} + T^{++}_{q,-\hat{x}}.\nonumber
\eea
For an arbitrary transverse spin direction 
the decomposition reads
\be
T^{++}_q = i\bar{q}\gamma^+
\stackrel{\leftrightarrow}{D^+}q
=\sum_{{\bf s}} \frac{i}{2}\bar{q}\left[ \gamma^+-s^j i\sigma^{+j}
\gamma_5\right]
\stackrel{\leftrightarrow}{D^+}
q =\frac{1}{2}\sum_{{\bf s}}\left[T^{++}_q + s^j \delta^jT^{++}_q\right]
=
\sum_{{\bf s}}T^{++}_{q,{\bf s}},
\ee
where the summation is over ${\bf s}=\pm(\cos \chi,\sin \chi)$, and
\be
\delta^jT^{++}_q \equiv \bar{q}\sigma^{+j}\gamma_5
\stackrel{\leftrightarrow}{D^+}q
\ee
represents the transversity asymmetry of the momentum density.
We will first present a heuristic argument for the transversity 
decomposition
of the angular momentum based on the
shift of parton distributions in impact parameter space.
However, for the sake of completeness, the heuristic derivation is followed by
a more formal derivation that 
parallels the approach chosen in Ji's original paper \cite{JiPRL}.

In the previous section we discussed that the 
total angular momentum carried by quarks of flavor $q$
can be associated with
the transverse shift of the center  of momentum of those quarks in a
target state that is described by a delocalized wave packet with
transverse polarization in the rest frame. Since $T^{++}_q$ does not
mix quark transversity, we can thus use this result to provide a
decomposition of the quark angular momentum into transversity 
eigenstates. 

For an unpolarized target, one might naively suspect that there is no
effect from the overall sideways shift of the transverse center 
of momentum discussed in Section \ref{sec:LC}. However, when one
considers the transversity asymmetry of the angular momentum, the
contributions from the two polarizations add up. As a result,
the transversity asymmetry in a delocalized target at rest contains
a term $\frac{1}{4} \int dx
H_T(x,0,0)\,x\,$. The parton model interpretation of this term is
the same as the term involving $H(x,0,0)$ in Ji's relation and it 
results from an overall transverse displacement of the center of
light-cone momentum in a state that is discribed by a delocalized
wavepacket centered around the origin in the rest frame.

In addition, there is the shift of the
transverse center of momentum arising from the deformation
in the center of momentum frame (\ref{Diehl}). 
Upon inserting Eq.(\ref{Diehl}) into (\ref{J}), and adding the
effect from the overall shift, 
we find for
the angular momentum carried by quarks with
transverse spin in the ${\bf s}$-direction in an unpolarized target
\be
\left\langle J_q^i({\bf s})\right\rangle
= \frac{1}{2}\left\langle J_q^i + s^j\delta^j J_q^i\right\rangle =
\frac{s^j}{2}
\left\langle \delta^j J_q^i\right\rangle = 
\frac{s^i}{4}
\int dx\left[ H_T(x,0,0) + 2\tilde{H}_T(x,0,0)+E_T(x,0,0)\right]x.
\label{cool}
\ee
The same GPDs that describe the distribution of transversity in
impact parameter space also characterize the correlation between
quark spin and angular momentum in an unpolarized target.

So far we have put special emphasis on drawing a connection
between the transverse distortion of impact parameter dependent
PDFs and the angular momentum of the quarks. In the following
we present an alternative derivation of Eq. (\ref{cool}) which
follows more the approach in Ref. \cite{JiPRL}. 
For this purpose we consider the 
form factor of the transversity
density with one derivative \cite{DH,DH2}
\bea
\left\langle p^\prime \right| \bar{q}\sigma^{\lambda \mu}
\gamma_5 i\!\stackrel{\leftrightarrow}{D^\nu} q \left| p
\right\rangle &=& \bar{u} \sigma^{\lambda \mu} \gamma_5 u\,
\bar{p}^\nu A_{T20}(t) + \frac{\varepsilon^{\lambda \mu \alpha
\beta} \Delta_\alpha \bar{p}_\beta \bar{p}^\nu}{M^2}\bar{u}u \,\tilde{A}_{T20}(t)
 + \frac{\varepsilon^{\lambda \mu \alpha
\beta} \Delta_\alpha  \bar{p}^\nu}{2M}
\bar{u}\gamma_\beta u \,B_{T20}(t)\nonumber\\
& &+ \frac{\varepsilon^{\lambda \mu \alpha
\beta} \bar{p}_\alpha \Delta^\nu}{M}\bar{u}\gamma_\beta u \,
\tilde{B}_{T21}(t) \label{formfactor}
\eea
where antisymmetrization in $\lambda$ and $\mu$ and
symmetrization in $\mu$ and $\nu$ is implied. 
The invariant
form factors in Eq. (\ref{formfactor}) are the second moments of
the chirally odd GPDs
\bea
A_{T20}(t) &=& \int_{-1}^1 dx x H_T(x,\xi,t) \label{second}\\
\tilde{A}_{T20}(t) &=& \int_{-1}^1 dx x \tilde{H}_T(x,\xi,t) 
\nonumber\\
B_{T20}(t) &=& \int_{-1}^1 dx x E_T(x,\xi,t) 
\nonumber\\
-2\xi \tilde{B}_{T21}(t) &=& \int_{-1}^1 dx x \tilde{E}_T(x,\xi,t) .
\nonumber\\
\eea
The projection operator on transverse spin (transversity)
eigenstates $P_{\pm \hat{x}} \equiv \frac{1}{2}\left( 1 \pm 
\gamma^x\gamma_5 \right)$ 
commutes with both $\gamma^0$, $\gamma^y$, and $\gamma^z$.
Hence neither $T_q^{0y}$ nor $T_q^{0z}$ mix between transversity
(in the $\hat{x}$ direction) and it is possible to decompose
\be
T_q^{0y} = T_{q,+\hat{x}}^{0y} + T_{q,-\hat{x}}^{0y} 
\ee
w.r.t. transversity, where
\be
T_{q,\pm\hat{x}}^{0y} = \frac{i}{2}\bar{q}\left[\gamma^0 D^y
+ \gamma^y D^0\right] P_{\pm \hat{x}} q
= \frac{1}{2}\left( T^{0y}_q \pm \delta^x T^{0y}_q\right)
\ee
where
\be
\delta^x T^{0y}_q = \frac{i}{2}\bar{q}\left(\gamma^0 
\stackrel{\leftrightarrow}{D^y} +
\gamma^y \stackrel{\leftrightarrow}{D^0}\right) \gamma^{x}\gamma_5 q
= -\frac{1}{2}\bar{q}\left( 
\sigma^{x0} \stackrel{\leftrightarrow}{D^y} 
+\,\sigma^{xy} \stackrel{\leftrightarrow}{D^0} \right) q
.
\ee
The same kind of decomposition can be made for $T_q^{0z}$. 
Evidently, these observations allow a similar decomposition for
\be
J^x_q = \int d^3x \left(yT^{0z}-zT^{0y}\right)
= J^x_{q,+\hat{x}}+ J^x_{q,-\hat{x}}.
\ee
The dependence of $J^x_q$ on the transversity of the quarks is
given by
\be
J^x_{q,\pm \hat{x}} = \frac{1}{2}\left( J^x_q
\pm \delta^x J^x_q\right)
\ee
where
\bea
\label{djJi}
\delta^x J^x_q = 
\int d^3x \left( \delta^x T^{0z}y -\delta^x T^{0y}z\right)=
\frac{1}{2}\int d^3x \,\bar{q}\left[
-\left( 
\sigma^{x0} \stackrel{\leftrightarrow}{D^z} 
+\,\sigma^{xz} \stackrel{\leftrightarrow}{D^0} \right)y
+
\left( 
\sigma^{x0} \stackrel{\leftrightarrow}{D^y} 
+\,\sigma^{xy} \stackrel{\leftrightarrow}{D^0} \right)z
\right]q .
\label{explicit}
\eea
The operators appearing in Eq. (\ref{explicit})
correspond to the operator appearing on the l.h.s. of
Eq. (\ref{formfactor}) with $\lambda = x$, $\mu=0$,
$\nu = z$,  and $\lambda = x$, $\mu=0$,
$\nu = y$ respectively. 
For the expectation value of the transversity asymmetry
$\delta^xJ^x_q = J^x_{q,+\hat{x}} - J^x_{q,-\hat{x}}$,  
Eq. (\ref{formfactor}) thus implies
in an unpolarized target at rest
\be
\left\langle
\delta^xJ^x_q \right\rangle
= \frac{1}{2}\left[A_{T20}+ 2\tilde{A}_{T20}(0) + B_{T20}(0)
\right], \label{djJi2}
\ee
which is the analogue of Ji's result 
$\left\langle J^i_q\right\rangle =  
S^i\left[A_{20}(0) + B_{20}(0) \right]$.
The angular momentum
$J^x$ carried by  quarks with transverse polarization
(transversity) in the $+\hat{x}$ direction in an unpolarized
target is one half of Eq. (\ref{djJi2}). 
Together with (\ref{second}) Eq. (\ref{djJi2}) provides an 
independent
confirmation of  the main result of this paper (\ref{cool}).
While the alternate derivation presented here is less intuitive 
than the light-cone approach, it serves to illustrate that the
result obtained in Sec. \ref{sec:LC} is gauge invariant
and independent of the light-cone framework.

While there exist several proposals to measure transversity 
$\delta q(x) = H_T(x,0,0)$,
it is not obvious how the other chirally odd GPDs which enter our relation
(\ref{cool}) can be
directly measured in an experiment. However, it should be straightforward to
determine these quantities in lattice QCD calculations 
\cite{latodd}, which
would provide valuable information about the correlation between 
angular momentum and spin of the quarks in an unpolarized target.
In addition, as we will discuss in the next section, the Boer-Mulders
effect may provide valuable information about the form factor entering
Eq. (\ref{djJi2}). Although this effect will not allow for a
quantitative experimental determination of the relevant moments
of chirally odd GPDs, it could provide useful information
on the sign and rough scale of these observables.
With this combined information it should be possible to
add another important piece of information to our
understanding of the spin structure of the nucleon.

\section{Boer-Mulders Effect}
In analogy to the Sivers effect, where quarks in a transversely
polarized target have a transverse momentum asymmetry which is 
perpendicular to the nucleon spin ${\bf S}$, 
it has been suggested that
there could also be an asymmetry of the transverse momentum of the 
quarks perpendicular to the quark spin ${\bf s}$
in an unpolarized target \cite{BM}
\bea
\mbox{Sivers:}\quad\quad f_{q/p^\uparrow}(x,\kT)&=&f_1^q(x,{\bf k}_\perp^2)
-f_{1T}^{\perp q}(x,{\bf k}_\perp^2)\frac{(\hat{\bf P}\times \kT)
\cdot{{\bf S}}}{M}
\label{BM}\\
\mbox{Boer-Mulders:}\quad\quad f_{q^\uparrow/p}(x,\kT)&=& \frac{1}{2}\left[
f_1^q(x,{\bf k}_\perp^2)
-h_{1}^{\perp q}(x,{\bf k}_\perp^2)\frac{(\hat{\bf P}\times \kT)
\cdot{{\bf s}}}{M}\right].
\eea
Here $f_{1T}^{\perp q}(x,{\bf k}_\perp^2)$ and 
$h_{1}^{\perp q}(x,{\bf k}_\perp^2)$ are referred to as the
Sivers and Boer-Mulders function respectively. Both the Sivers as
well as the Boer-Mulders function require a nontrivial FSI. 
In Refs. \cite{mb2} it has been suggested that
the transverse distortion of impact parameter dependent (unpolarized) 
quark distributions in a transversely polarized target can give rise
to a Sivers effect. If the quarks before they are being knocked out 
of the nucleon in SIDIS have a preferential direction in position 
space then the FSI can translate this position space asymmetry into
a momentum space asymmetry. Since the FSI is expected 
to be attractive on 
average, this means that a transverse distortion in the $+\hat{x}$ 
direction would translate into a momentum asymmetry in the 
$-\hat{x}$ direction. 

The distortion in impact parameter space for quarks with
flavor $q$ can be related to $\kappa_q$, i.e. the contribution to the
anomalous magnetic moment (with the electric charge of the quarks 
factored out) from the same quark flavor \cite{gpde}.  Within the
heuristic mechanism for the Sivers effect developed in Refs. 
\cite{mb2,mb3} one thus finds that the average Sivers effect for 
flavor $q$ and $\kappa^q$ should have opposite signs
\be
f_{1T}^{\perp q} \sim -\kappa^q .
\label{fkappa}
\ee
The signs for the predicted Sivers effect for $u$ and $d$ quarks in 
a proton have recently been confirmed by the HERMES collaboration
\cite{hermes}. Furthermore, the correlation above (\ref{fkappa}) has
been observed in a number of toy model calculations as well \cite{model}.


As far as the transverse distortion of transverse polarized quark
distributions is concerned, the forward matrix element of
$2\tilde{H}_T+E_T$, i.e.
\be
\kappa_T^q\equiv \int dx \left[ 2\tilde{H}_T(x,0,0)+E_T(x,0,0)
\right]
\ee
plays a role similar to the anomalous magnetic moment $\kappa^q$
for the unpolarized quark distributions on a transverse polarized 
target. Indeed, $\kappa_T^q$ governs the transverse spin-flavor dipole
moment in an unpolarized target (\ref{spindipole}).
Indeed, $\kappa_T^q$ tells us, in units of $\frac{1}{2M}$,
how far and in which direction 
the average position of quarks with spin in the $\hat{x}$
direction, is shifted in the $\hat{y}$ direction for an unpolarized
target relative to the
transverse center of momentum.

Encouraged by the success of the impact parameter distortion based
mechanism for the Sivers effect, we propose a similar semi-classical
mechanism for the Boer-Mulders effect: if $\kappa_T>0$, then the
distribution for quarks polarized in the $+\hat{y}$ direction is
shifted towards the $-\hat{x}$ direction (Fig. \ref{wirbel}). The 
FSI is expected  to have a qualitatively similar effect on deflecting
this distorted position space into the opposite direction, i.e.
for $\kappa_T>0$ we expect that quarks  polarized in the $+\hat{y}$ 
direction should be preferentially deflected in the $+\hat{x}$
direction. In accordance with the Trento convention (\ref{BM})
\cite{Trento} this implies that $h_{1}^{\perp q}<0$. More generally,
we expect that on average the Boer-Mulders function for 
flavor $q$ and $\kappa^q_T$ should have opposite signs
\be
h_{1}^{\perp q} \sim -\kappa^q_T .
\label{hkappa}
\ee
In appendix \ref{correl} some of the arguments from Ref. \cite{mb3}
are repeated for the case of $h_{1}^{\perp q}$. For a more detailed
discussion the reader is refered to Refs. \cite{mb2,mb3}.

Furthermore, up to a rescaling by the factor $\kappa^q_T/\kappa^q$,
we expect the average Boer Mulders function to be of roughly 
the same scale as the Sivers function.

In the case of the transverse distortion of chirally even impact 
parameter dependent parton distributions, the quantity that determines
the magnitude of the distortion, i.e. the anomalous magnetic
moment $\kappa^q$, is known experimentally (up to uncertainties
from the contribution of $s$ quarks). In the chirally odd case
essentially nothing is known about the corresponding quantity
$2\tilde{H}_T(x,0,0)+E_T(x,0,0)$ from experiment, although the long 
distance tail
of chirally odd GPDs might be accessible in diffractive
electroproduction of vector meson pairs \cite{pire}. 
Therefore it would be very useful
to determine this quantity in lattcie QCD, so that at least a 
rough estimate can be made for sign and magnitude of the 
Boer-Mulders function.

\section{Summary}
We have studied the light-cone momentum density of a
delocalized, but axially symmetric wave packet describing a 
transversely polarized particle that is at rest.
Two effects lead to deviations from axial symmetry in the 
resulting momentum density. For a particle with a nontrivial
internal structure (e.g. if it has an anomalous magnetic moment)
there is an intrinsic asymmetry,
relative to the particle's center of momentum, that is 
described by the GPD $E(x,0,0)$.
In addition, the center of momentum of the whole wave packet
is shifted sideways relative to the center of instant form
wave packet by half a Compton wavelength. This sideways shift is
responsible for the term proportional to $xH(x,0,0)$ in Ji's 
relation.

The $T^{++}$ component of the energy momentum tensor that appears
in the angular momentum relation does not mix transversity and
it is therefore possible to decompose the angular momentum into
transversity components. The information to be gained by performing
this decomposition is the correlation between the transverse spin
and the transverse angular momentum carried by the quarks.
We find that the correlation between transverse spin  and
angular momentum of the quarks in an unpolarized target
is described by a linear combination of chirally odd 
GPDs $H_T(x,0,0)+2\tilde{H}_T(x,0,0)+E_T(x,0,0)$.

The same linear combination of GPDs  ($2\tilde{H}_T(x,0,0)+E_T(x,0,0)$)
that appears in the 
correlation between transverse spin  and
angular momentum of the quarks in an unpolarized target
also describes the transverse displacement of quarks with a
given transversity in an unpolarized target relative to the center
of momentum. 
We suggest that the resulting 
angular dependence of the chirality density, in 
combination with the final state interaction, gives rise to  the
T-odd Boer-Mulders effect and we make a prediction for the
sign of the Boer-Mulders effect in terms of those GPDs.
Lattice determinations of chirally
odd GPDs can thus be used to predict the sign of the Boer-Mulders
function. Likewise, an experimental measurement of the Boer-Mulders
function could be used to learn about the correlation between
transverse spin  and angular momentum of the quarks in an 
unpolarized target.

{\bf Acknowledgements:}
I would like to thank 
M.Anselmino, A.Bacchetta, M.Diehl, P.H\"agler,
X.Ji, A.Metz, P.Muilders, F.Pijlman, 
and G.Schnell for useful discussions.
This work was partially supported by the DOE under grant number 
DE-FG03-95ER40965.
\appendix
\section{Angular Momentum and Light-Cone Momentum Density}
The ansatz for the form factor of the energy momentum tensor
(\ref{T}) implicitly uses of Lorentz invariance. In this appendix,
we will demonstrate how these symmetries enter the derivation
of the representation of the angular momentum in terms of the
impact parameter dependent light-cone momentum density $T^{++}$.

We start from the expectation value of 
$J^x_q=T_q^{0z}y-T_q^{0y}z$ taken in a delocalized
wave packet at rest. We will furthermore consider the specific
example where the target is an eigenstate of the total angular 
momentum in the $\hat{x}$-direction, i.e. up to a phase, it is
invariant under rotations about the $\hat{x}$ axis.
Upon performing a $90^o$-rotation around the $\hat{x}$ axis
one thus finds
\be
\int d^3{\bf r} \left\langle T_q^{0y}({\bf r})\right\rangle z
=-\int d^3{\bf r} \left\langle T_q^{0z}({\bf r})\right\rangle y
\ee
and therefore
\be
\left\langle J^x_q\right\rangle = 2
\int d^3{\bf r} \left\langle T_q^{0z}({\bf r})\right\rangle y .
= \int d^3{\bf r} \left\langle \left[
T_q^{0z}({\bf r})+T_q^{z0}({\bf r})\right]\right\rangle y ,
.
\label{T0z}
\ee
Likewise, performing a $180^o$ rotation around the 
$\hat{x}$ axis yields
\bea
\int d^3{\bf r} \left\langle T_q^{00}({\bf r})\right\rangle y
&=&-\int d^3{\bf r} \left\langle T_q^{00}({\bf r})\right\rangle y
=0\\
\int d^3{\bf r} \left\langle T_q^{zz}({\bf r})\right\rangle y
&=&-\int d^3{\bf r} \left\langle T_q^{zz}({\bf r})\right\rangle y
=0\nonumber
\eea
and therefore $T_q^{0z}+T_q^{z0}$ in Eq. (\ref{T0z}) can be 
replaced by $2T^{++}_q =  T_q^{0z}+T_q^{z0}+T_q^{00}+T_q^{zz}$
yielding
\be
\left\langle J^x_q\right\rangle = 2 \int
d^3{\bf r} \left\langle T_q^{++}({\bf r})\right\rangle y.
\ee
Finally, for a delocalized wave packet describing a state with
zero momentum, $T_q^{++}({\bf r}$ is time-independent. 
It is thus possible to 
replace $\sqrt{2}\int dz \longrightarrow \int d x^-$ in these
integrals, yielding a light-cone representation of the angular
momentum in terms of twist-2 operators
\be
\left\langle J^x_q\right\rangle =\sqrt{2}
\int dx^- \int d^2 {\bf r}_\perp 
\left\langle T_q^{++}({\bf r})\right\rangle y
\ee
in agreement with Eqs. (\ref{T++AB},\ref{J}).
\section{Quark Correlations and the Boer-Mulders Function}
\label{correl}
In this appendix, we will follow the approach in Ref. \cite{mb3}
and relate the Boer-Mulders (BM) function to color density-density
correlations in the transverse plane. The gauge invariant 
operator definition of the unintegrated transversity density
relevant for the BM function in SIDIS reads
\be
\delta^i q(x,\kT) \equiv \int \frac{dy^-d^2\yT}{16\pi^3}
e^{-ixp^+y^-+i\kT\cdot \yT}
\left\langle p \left| \bar{q}_U(y) U_{[\infty^-,\yT;\infty^-,\0T ]}
\sigma^{+i}\gamma_5 q_U(0)\right|p\right\rangle,
\label{opdef}
\ee
with
\bea
q_U(0)&=&U_{[\infty^-,\0T;0^-,\0T ]}q(0)\\
\bar{q}_U(y)&=& \bar{q}(y) U_{[y^-,\yT;\infty^-,\yT ]}.
\eea
The $U$'s are Wilson line gauge link, for example
\be
U_{[0;\xi ]} = P\exp\left(ig\int_0^1 ds \xi_\mu A^\mu(x\xi)\right)
\ee
connecting the points $0$ and $\xi$. The choice of paths in Eq. 
(\ref{opdef}) is not arbitrary, but reflects the final state 
interactions, as the ejected quark travels along the light-cone.
The gauge link segment at light-cone infinity is formally
necessary to render Eq. (\ref{opdef}) gauge invariant, but plays
an important role only in the light-cone gauge $A^+=0$, where the
segments along the light-cone do not contribute.

Following Ref. \cite{mb3}, we evaluate (\ref{opdef}) in the 
light-cone gauge, yielding for the average transverse momentum
\be
\int d^2\kT \delta^i
q(x,\kT) \kT
= -g \int \frac{dy^-}{4\pi}e^{-ixp^+y^-}
\left\langle p \left| \bar{q}(y^-,\0T) \frac{\lambda^a}{2}\sigma^{+i}\gamma_5 q 
{\bf A}_\perp^a (\infty^-,\0T)
\right|p\right\rangle,\label{av1}
\ee
where $\lambda^a$ are the Gell-Mann matrices.
In Ref. \cite{mb3}, a constraint condition on the gauge
field at $y^-=\infty$ were derived. Solving those to lowest order
and inserting the result back into Eq. (\ref{av1}) yields
\be
\int d^2\kT \delta^i
q(x,\kT) k_\perp^j = 
-\frac{g}{2}\int \frac{dy^-}{4\pi}e^{-ixp^+y^-}
\int \frac{d^2\xT}{2\pi} \left\langle p \right| \bar{q}(y^-,\0T)
\frac{\lambda^a}{2}\sigma^{+i}\gamma_5 q(0) \rho^a(\xT)\left|p\right\rangle
\frac{x^j}{{\bf x}_\perp^2}
\label{av2}
\ee
where
\be
\rho^a(\xT) = g\int dx^- \left[ -gf^{abc}A_i^b\partial_-A_i^c+ 
\sum_q \bar{q}\gamma^+ \frac{\lambda^a}{2}q\right]
\ee
is the color charge density at position $\xT$ integrated over all
$x^-$. The average transverse momentum can thus be related to the
transverse color density-density dipole-correlations between the
transversity density and the spin averaged density of all partons.

Eq. (\ref{av2}) is equivalent to treating the FSI in first order
perturbation theory, and can, to this order, be used to justify
our intuitive picture developed above. If we study the asymmetry at 
a relatively low $Q^2$ scale and/or large $x$, the correlation in 
Eq. (\ref{av2}) should be dominated by the valence quarks where
the Gell-Mann matrices can effectively be replaced by an overall 
color factor, which is negative due to the attractive nature of the 
QCD  potential in a color singlet nucleon. Therefore, we can relate
the average transverse momentum to the color neutral density-density
correlation. 
If the chirally odd GPDs exhibit a large transverse dipole
moment ($\kappa_T$ large) it almost impossible not to get a
significant
density-density dipole-correlations between the
transversity density and the spin averaged density of all partons.
This is true even though we do not expect the density-density
correlation to factorize.
The rest of the argument is the same as in Ref. \cite{mb3}.
Nonperturbatively,
both for $f_{1T}^\perp$ \cite{mb3} as well as for $h_1^\perp$, we 
are guided more by intuition in order to arrive at 
Eqs. (\ref{fkappa},\ref{hkappa}).

However, there is one important difference between $f_{1T}^\perp$ and
$h_1^\perp$: in the case of $f_{1T}^\perp$ we were able to show
that the net Sivers effect, summed over all flavors and the glue and
integrated over all momenta must vanish \cite{mbg}. No such 
statement can be made for $h_1^\perp$.

\bibliography{dist_resub.bbl}
\end{document}